\documentclass[final]{elsarticle}

\usepackage{lineno,hyperref}
\usepackage[table]{xcolor}
\usepackage{amsmath,amsfonts,amsthm,amssymb}
\usepackage{graphicx}
\usepackage[margin=1in]{geometry}
\usepackage[section]{placeins}
\usepackage{subcaption}
\usepackage{gensymb}
\graphicspath{{figures/}}
\usepackage[separate-uncertainty,table-align-uncertainty,multi-part-units=repeat]{siunitx}

\usepackage{xargs}
\journal{Journal of Cardiovascular Computed Tomography}

\begin{document}

\begin{frontmatter}

\title{Sex-Specific Variances in Anatomy and Blood Flow of the Left Main Coronary Bifurcation: Implications for Coronary Artery Disease Risk}
\author[inst1]{Ramtin Gharleghi\corref{cor1}}
\ead{r.gharleghi@unsw.edu.au}
\affiliation[inst1]{organization={School of Mechanical and Manufacturing Engineering, UNSW},
            city={Sydney},
            postcode={2052}, 
            state={NSW},
            country={Australia}}
\author[inst1]{Mingzi Zhang}
\author[inst2,inst3]{Dona Adikari}
\author[inst4]{Lucy McGrath-Cadell}
\author[inst4]{Robert M. Graham}
\author[inst5]{Jolanda Wentzel}
\author[inst6]{Mark Webster}
\author[inst6]{Chris Ellis}
\author[inst2,inst3]{Sze-Yuan Ooi}
\author[inst1]{Susann Beier}
\affiliation[inst2]{organization={Prince of Wales Clinical School of Medicine},
            city={Sydney},
            postcode={2053}, 
            state={NSW},
            country={Australia}}
\affiliation[inst3]{organization={Department of Cardiology, Prince of Wales Hospital},
            city={Sydney},
            postcode={2053}, 
            state={NSW},
            country={Australia}}
\affiliation[inst4]{organization={Molecular Cardiology and Biophysics Division, Victor Chang Cardiac Research Institute},
            city={Sydney},
            country={Australia}}
\affiliation[inst5]{organization={Erasmus},
            city={},
            country={Netherlands}}
\affiliation[inst6]{organization={ Auckland City Hospital},
            city={Auckland},
            country={New Zealand}}

\cortext[cor1]{The authors declare that they have no conflicts of interest\\Corresponding Author, Ainsworth Building, High St., Sydney, NSW Australia\\\hphantom~~~~~Phone No: +612 9385 4097}

\begin{abstract}

\textbf{Background:} Studies have shown marked sex disparities in Coronary Artery Diseases (CAD) epidemiology, yet the underlying mechanisms remain unclear. We explored sex disparities in the coronary anatomy and the resulting haemodynamics in patients with suspected, but no significant epicardial coronary artery disease.

\textbf{Methods:} Left Main (LM) coronary artery bifurcations were reconstructed from Computed Tomography Coronary Angiography (CTCA) images of 127 cases (42 males and 85 females, aged 38 to 81). Detailed 3D-anatomical shape parameters were measured for comparison, including bifurcation angles, curvature representing tortuosity, and diameters, before solving the clinically relevant haemodynamic metrics using Computational Fluid Dynamics (CFD). The severity and location of the normalised vascular area exposed to physiologically adverse haemodynamics  were statistically compared between sexes for all LM branches, i.e., the LM, Left Anterior Descending (LAD), and Left Circumflex (LCx), while accounting for different sample sizes.

\textbf{Results: }
We found significant differences between sexes in potentially adverse haemodynamics. Interestingly, females were more likely than males to exhibit adversely low Time Averaged Endothelial Shear Stress TAESS along the inner wall of a bifurcation (16.8\% vs. 10.7\%). Males had a higher percentage of areas exposed to both adversely high Relative Residence Time RRT (6.1\% vs 4.2\%, p=0.001) and high Oscillatory Shear Index OSI (4.6\% vs 2.3\%, p\textless0.001). However, the OSI values were generally small and should be interpreted cautiously. These haemodynamic differences were compounded by anatomical differences, whereby males had larger arteries (M vs F, LM: 4.0mm vs 3.3mm, LAD: 3.6mm 3.0mm, LCX:3.5mm vs 2.9mm), and females exhibited higher curvatures in all three branches (M vs F, LM: 0.40 vs 0.46, LAD: 0.45 vs 0.51, LCx: 0.47 vs 0.55, p\textless0.001) and larger inflow angle of the LM trunk (M: 12.9\degree vs F: 18.5\degree, p=0.025). There was no difference in bifurcation angles.

\textbf{Conclusions:} Haemodynamic differences were found between male and female patients, which may contribute, at least in part, to differences in CAD risk between sexes. This work may facilitate a better understanding of sex differences in the clinical presentation of CAD and other causes of myocardial ischemia, ultimately contributing to improved sex-specific screening and therapeutic strategies and, thus, improved outcomes, especially relevant for women with CAD who currently have worse predictive outcomes. 
\end{abstract}


\end{frontmatter}

\section{Introduction}
Sex disparities in the incidence rates, clinical presentation, and management of CAD are well recognised. Despite a large and growing body of research into these disparities, women fared worse in most treatment metrics \cite{mnatzaganian_SexDisparitiesAssessment_2020}. Previous scientific efforts towards rectifying these inequalities have predominantly focused on the differences in symptoms, physiology, and medical care provided \cite{presbitero_GenderDifferencesOutcome_2003}. Even though the link between atherosclerotic plaque development and coronary anatomy is well established \cite{eshtehardi2012association, morbiducci_AtherosclerosisArterialBifurcations_2016}, few studies have investigated sex differences in coronary anatomy and haemodynamics. 

Established anatomical differences are that curvature/tortuosity \cite{kashyap2022accuracy} is higher in females \cite{sharma2019risk,ciurica_ArterialTortuosity_2019,kahe2020coronary}, and that coronary artery diameters are larger in males even after correcting for body size \cite{ilayperuma_SexualDifferencesDiameter_2011,chiha_GenderDifferencesPrevalence_2017}. These findings raise the question: Could an anatomical difference between female and male CAD patients contribute to different haemodynamic profiles, leading to potential gender differences in the propensity to develop CAD?

Coronary anatomy is the predominant factor determining the local haemodynamics to which endothelial cells are exposed. Such blood-flow-induced haemodynamic shear stresses on the inner arterial cell lining are a well-established driver of adverse biological cellular response \cite{fisher2001endothelial}. Specifically, studies revealed that endothelial cells exposed to extremely low Endothelial Shear Stress (ESS, \textless 0.5 Pa) are prone to increased tissue growth factor secretion, promoting neointimal thickening and plaque progression \cite{cecchi2011role, meyerson2001effects}. Moreover, frequent changes in the local ESS direction, quantified as the Oscillatory Shear Index (OSI, \textgreater 0.1) \cite{xie2014computation}, were found to be associated with atheroma formation \cite{ku_PulsatileFlowAtherosclerosis_1985}. The Relative Residence Time (RRT) has been proposed \cite{himburg2004spatial} to capture low and fluctuating ESS with RRT \textgreater 4.17 Pa$^{-1}$ considered adverse \cite{rabbi2020computational,chiastra_HealthyDiseasedCoronary_2017}.

Despite the well-established understanding that anatomical characteristics induce clinically adverse haemodynamic endothelial cell response, the differences between males and females remain under-explored. Existing literature has reported smaller \cite{ilayperuma_SexualDifferencesDiameter_2011} and more  \cite{eleid_CoronaryArteryTortuosity_2014} coronary arteries in women, as well as a larger inflow angle \cite{medrano2016computational} between the LM and the bifurcation plane. These studies did not analyse the impact of these differences on the resulting haemodynamics. To our knowledge one study \cite{wentzel2022sex} considered haemodynamic differences, finding similar ESS between men and women, however this study only considered differences in coronary diameter. The haemodynamic differences resulting from different coronary anatomy is expected to affect both the risk and presentation of CAD \cite{taqueti2018sex}, therefore necessitating more a detailed analysis of these haemodynamic conditions. Following our previous study in which we systematically analysed coronary anatomy in a large cohort of patients with suspected CAD \cite{medrano-gracia_ConstructionCoronaryAtlas_2014,medrano-gracia_StudyCoronaryBifurcation_2016}, we now explore the haemodynamic differences in males and females, thereby identifying potential sex-specific risk factors.

\section{Methods}

\subsection{Study Cohort}
This study was approved by the local institutional ethics committees of the University of Auckland (Ref. 022961) and the University of New South Wales UNSW (Ref. HC190145), with informed written consent received from all participants. Patients were selected from those referred for imaging to investigate cardiovascular disease. We excluded patients with (i) reported atherosclerotic plaques or non-zero calcium score, and (ii) CTCA images of insufficient quality for segmentation and thus computational modelling, for example, due to respiratory or motion artefacts. A total of 127 CTCAs (42 men and 85 women), were included. The patient demographics are shown in Table \ref{tab:cfd-demos}.

\begin{table}[]
\centering
\caption{Summary of the 127 patient demographics by sex.}
\label{tab:cfd-demos}
\rowcolors{2}{gray!25}{white}
\begin{tabular}{l|c|c}
                & \textbf{Male}           & \textbf{Female}         \\\hline
\textbf{Ethnicity}       & Caucasian 86\%/ Other 14\% & Caucasian 88\%/ Other 12\% \\
\textbf{Weight {[}kg{]}*} & 88.4\textpm11.4   & 72.7\textpm12.9\\
\textbf{Height {[}m{]}*}  & 1.78\textpm0.075  & 1.63\textpm 0.067\\
\textbf{Body Mass Index (BMI)}             & 27.8\textpm3.55 & 27.2\textpm 4.72   \\
\textbf{Smoking history }& 26.1\% (11)    & 35.3\% (30)    \\
\textbf{Hypertension}    & 28.6\% (12)    & 28.2\% (24)   \\
\textbf{Diabetes}        & 0\%  (0)          & 3.5\% (3)         \\
\textbf{Age {[}years{]}*} & 53.3 \textpm6.18 & 57.2\textpm7.67   \\
\multicolumn{3}{l}{\scriptsize{*mean value was statistically different between groups}}
\end{tabular}
\end{table}

\subsection{CTCA Acquisition Protocol}
All patients underwent CTCA scans using a 64-slice CT scanner (GE LightSpeed, USA) with retrospective ECG-gated acquisition. The contrast-enhanced images were obtained with intravenously administered 60-80 mL of a non-ionic medium (Omnipaque 350, GE Healthcare, USA). Where necessary, beta-blockers were administered to reduce heart rate to around 60 beats per minute. Images were taken at the end-diastolic phase for each patient.

\subsection{Vessel reconstruction and shape analysis}
\begin{figure}[h]
    \centering
    \includegraphics[width=0.75\textwidth]{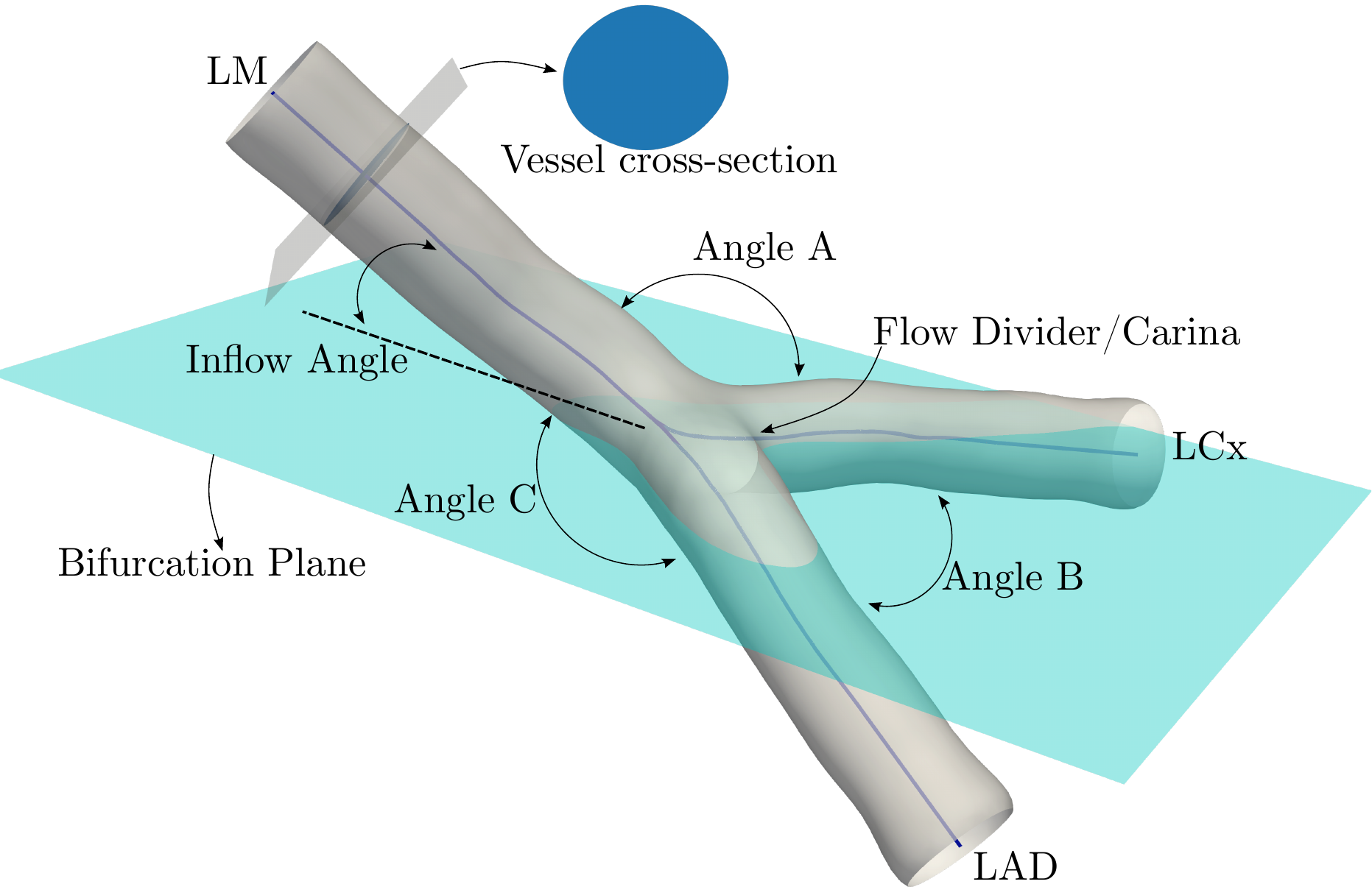}
    \caption{Left main coronary shape definition: Anatomical parameters defined for the Coronary Left Main Bifurcation comprised of the Left Main (LM), Left Anterior Descending (LAD), and Left Circumflex (LCx) branch.}
    \label{fig:parameters}
\end{figure}

The coronary arteries were segmented and reconstructed from the patients' CTCA images, as previously reported \cite{medrano-gracia_StudyCoronaryBifurcation_2016}. OsiriX (v4.1.2) was used for semi-automatic segmentation of the arteries, producing a triangulated surface mesh. The left main bifurcation was extracted from the coronary tree at a maximum arc length of 10mm from the bifurcation point. The bifurcation was smoothed using a Taubin filter \cite{taubin_CurveSurfaceSmoothing_1995} with a passband of 0.03 and 30 iterations. The shape parameters were measured using an in-house python script and included the inflow angle, and bifurcation angles A, B and C as defined by the European Bifurcation Club \cite{ormiston2018bench} (Fig. \ref{fig:parameters}).

Angles were measured between the average tangent vectors for the first 5mm of the vessels. The inflow angle is the angle between LM and the bifurcation plane \cite{medrano-gracia_StudyCoronaryBifurcation_2016}, defined as the best-fit plane containing the first 5mm of LAD and LCx centerlines. The median diameter for each vessel, corresponding to the diameter of a circle with equivalent cross-sectional area, was calculated over the 10mm segment comprised of intervals with 0.01 mm spacing. As the change in vessel diameter along the vessel length is relatively small (standard deviation of 0.1 mm for calculated diameters), the median should be a representative measure. Finet's ratio (FR) is defined as the ratio of the LM diameter over the sum of the daughter branch diameters ($d_{LM}/(d_{LAD}+d_{LCx})$) \cite{finet2008fractal,ormiston2018bench}. We calculated the mean curvature of all branches, which was demonstrated as a superior tortuosity measure than the commonly used tortuosity index \cite{kashyap2022accuracy}. 

\subsection{Haemodynamic Simulation}
Following a sensitivity test, we generated unstructured tetrahedral meshes using TetGen \cite{si_TetGenDelaunaybasedQuality_2015} with an average of 2 million elements per bifurcation (average element volume of 1.77$\cdot$10$^{-4}$ \si{mm^3} $\pm$ 1.43$\cdot$10$^{-4}$ \si{mm^3}). The timestep size was 0.002 seconds. The mesh was imported into ANSYS CFX v19.3 (Canonsburg, PA) to resolve the haemodynamic metrics. A generic flowrate waveform \cite{nichols_CoronaryCirculation_2011} was prescribed at the extended inlet, scaled based on the relationship (flowrate = \(1.43\cdot d_{LM}^{2.55}\))  as proposed by Giessen et al.\cite{vandergiessen_InfluenceBoundaryConditions_2011}. A plug flow profile was applied as suitable to simulate a region close to the coronary ostia. A constant reference pressure of 0Pa was specified at the outlets appropriate for non-diseased vessels, allowing for a computationally stable solution and validated \textit{in vitro} \cite{beier2016dynamically}. The vessels were modelled as rigid in line with \cite{eslami2020effect}. A laminar non-Newtonian Carreau-Yasuda \cite{razavi2011numerical} fluid model was prescribed, resulting in average Reynolds numbers of 600. The simulations were run on the Katana computing cluster \cite{katana}.

\subsection{Haemodynamic Conditions}
Following previous studies \cite{malek_HemodynamicShearStress_1999,morbiducci_AtherosclerosisArterialBifurcations_2016}, we considered the local haemodynamic environment adverse to the endothelial cells if coronary arteries are exposed to (i) Low Time Averaged Endothelial Shear Stress (also commonly referred to as Wall Shear Stress WSS in previous literature), i.e. TAESS \textless 0.5 Pa, (ii) High Oscillatory Shear Index, i.e. OSI \textgreater 0.1, or (iii) high Relative Residence Time, i.e., RRT \textgreater 4.17 Pa $^{-1}$. We extracted the ESS vector $\mathbf{\tau_w}$ from the fourth cardiac cycle to minimise the transient start-up effects. The TAESS, OSI, and RRT were then derived as follows:
\begin{align}
    \text{TAESS} &= \frac 1 T \int_0^T|\mathbf{\tau_w}|dt \\
    \text{OSI} &= \frac 1 2 \left(1- \frac{|\int_0^T \mathbf{\tau_w} dt|}{\int_0^T |\mathbf{\tau_w}| dt}\right)\\
    \text{RRT} &= \frac{1}{(1-2\cdot\text{OSI})\cdot\text{TAESS}}
\end{align}
where $T$ is the cardiac cycle period. As well as analysing the normalised adverse area size of these measurements, we also studied their respective locations within the left main branches. The percentage of the vessel area exposed to low TAESS, high OSI, and high RRT are referred to as \%lowTAESS@0.5, \%highOSI@0.1, and \%highRRT@4.17, respectively.

\subsection{Statistical Modelling}
Continuous variables describing the study cohort are presented as mean ± Standard Deviation (SD). Nominal variables are presented as numbers and percentages. Sex was determined as significant using a type-I Analysis of Variance (ANOVA) test before using a Welch’s t-test to determine the difference of means for each shape parameter, i.e. diameter, curvature, bifurcation angles A, B, C and Inflow. Welch’s t-test was used for statistical comparisons as it is robust to the different variances of the two groups \cite{derrick2016welch}, and normality violations \cite{delacre2017psychologists}. Body size is a confounding factor for arterial diameter \cite{friedman_ArterialGeometryAffects_1983}, which was accounted for using both the Body Surface Area (BSA), and Body Mass Index (BMI) in a multivariate model. The results were analysed using python and its statsmodels package (v0.11.0). A standard linear model assumption was used with p\textless0.05 considered statistically significant. Statistically significant differences are marked with an asterisk$^*$.

\section{Results}
\subsection{Geometric differences between males and females}
\begin{figure}[h]
    \centering
    \includegraphics[width=0.9\textwidth]{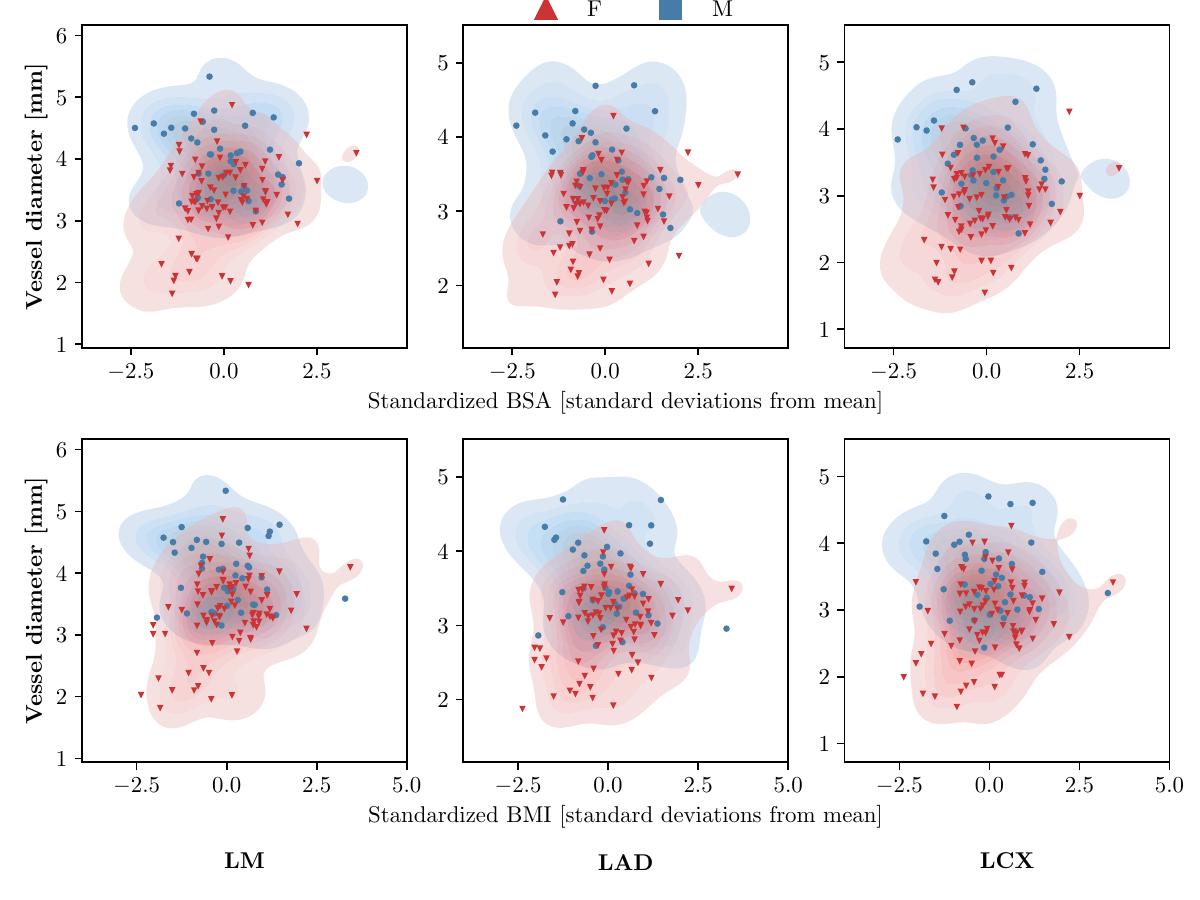}
    \caption{Distribution of left main branch diameters by sex: Vessel diameter of Left Main (LM), Left Anterior Descending (LAD), and Left Circumflex (LCx, left to right) for females (red) and males (blue) 
over standardised BSA (top) and BMI (bottom).}
    \label{fig:kdes}
\end{figure}

Sex-specific differences in the curvature and diameter of all branch vessels were found, see Table \ref{tbl:params}.  Specifically, diameters of the coronary arteries were larger (all p \textless 0.001*) in males (LM: \SI{4.02(0.52)}{}, LAD: \SI{3.62(0.50)}{}, and LCx: \SI{3.51(0.53)}{}) than in females (LM: \SI{3.33(0.61)}{}, LAD: \SI{3.01(0.50)}{}, and LCx: \SI{2.88(0.57)}{}), as shown in Fig. \ref{fig:kdes}. The differences in vessel diameter remained statistically significant even after adjusting for BMI (p\textless0.001* for all branches), and BSA (p=0.001* for all branches) by including these in a multivariate model.

\begin{table}[h]
\centering
\caption{Anatomical characteristics by sex and overall.}
\label{tbl:params}
\rowcolors{2}{gray!25}{white}
\begin{tabular}{lS[table-format=3.3(3)]S[table-format=3.3(3)]S[table-format=2.3]cc}\hline
\textbf{Feature} & {\textbf{Male}} & \textbf{Female} & \textbf{p-value} & \textbf{Difference} & \textbf{95\% CI} \\
&&&&\textbf{M – F} &\\\hline
\textbf{Angle A [$^{\circ}$]}      & 125.4(17.1)  & 122.9(21.2)    & 0.516            & 2.47   & {[}-4.54  9.47{]}   \\
\textbf{Angle B [$^{\circ}$]}      & 81.9(18.6)   & 77.9(18.6)   & 0.267            & 3.96   & {[}-3.10  11.0{]}    \\
\textbf{Angle C [$^{\circ}$]}      & 146.7(10.7)  & 150.1(21.7)  & 0.244            & -3.39  & {[}-9.12  2.34{]}   \\
\textbf{Inflow Angle [$^{\circ}$]} & 12.9(11.2)   & 18.5(15.9)   & 0.025*            & -5.57  & {[}-10.4  -0.70{]} \\
\textbf{LM Diameter [mm]}   & 4.02(0.52)   & 3.33(0.61)   & \textless0.001* & 0.69   & {[}0.48 0.90{]}     \\
\textbf{LAD Diameter [mm]}  & 3.62(0.50)    & 3.01(0.50)    & \textless0.001* & 0.61   & {[}0.42 0.79{]}     \\
\textbf{LCx Diameter [mm]}  & 3.51(0.53)   & 2.88(0.57)   & \textless0.001* & 0.63   & {[}0.42 0.84{]}     \\
\textbf{LM Curvature [mm$^{-1}$]} & 0.40(0.08) & 0.46(0.12) & 0.001* & -0.062& {[}-0.01 -0.03{]} \\
\textbf{LAD Curvature [mm$^{-1}$]}&  0.45(0.07) & 0.51(0.11) & 0.001* & -0.063	& {[}-0.10 -0.03{]} \\
\textbf{LCx Curvature [mm$^{-1}$]}&  0.47(0.08) &0.55(0.13)  & 0.001* & -0.072& {[}-0.11 -0.03{]} \\
\textbf{Finet's ratio}      & 0.57(0.02)   & 0.57(0.04)   & 0.984            & 0.0001 & {[}-0.01  0.01{]}  
\end{tabular}

\end{table}

\begin{figure}
\centering
\includegraphics[width=0.6\linewidth]{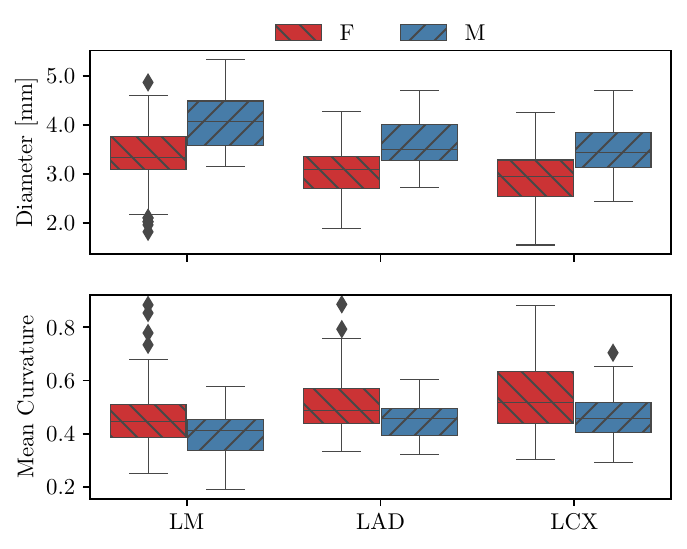}
\caption{Sex differences in left main curvature and branch diameters: Left main branch diameters (top), and mean curvature (bottom) for females (red) and males (blue) in the Left Main (LM), Left Anterior Descending (LAD), and Left Circumflex (LCx). The diameter (p\textless0.001*), and mean curvature (p\textless 0.01*).}.
\label{fig:diamtort}
\end{figure}

The mean curvature was significantly higher in female patients for all branches (M vs F, LM: \SI{0.40(0.08)}{} vs \SI{ 0.46(0.12)}{}, LAD: \SI{0.45(0.07)}{} vs \SI{0.51(0.11)}{}, LCX: \SI{0.47(0.08)}{} vs \SI{0.55(0.13)}{}, p=0.001*). Curvature and diameter sex-differences are visualised in Fig. \ref{fig:diamtort}. Of the bifurcation angles, only the inflow angle varied between males and females, with females showing slightly larger acute inflow angles than males (M vs F: \SI{12.9(11.2)}{\degree} vs \SI{18.5(15.9)}{\degree}, p = 0.025*). No statistical difference in the Finet's Ratio was found. 
All results are listed in Table \ref{tbl:params}.

\FloatBarrier

\subsection{Hemodynamic Differences}
\begin{figure}[h]
\centering
\includegraphics[width=\linewidth]{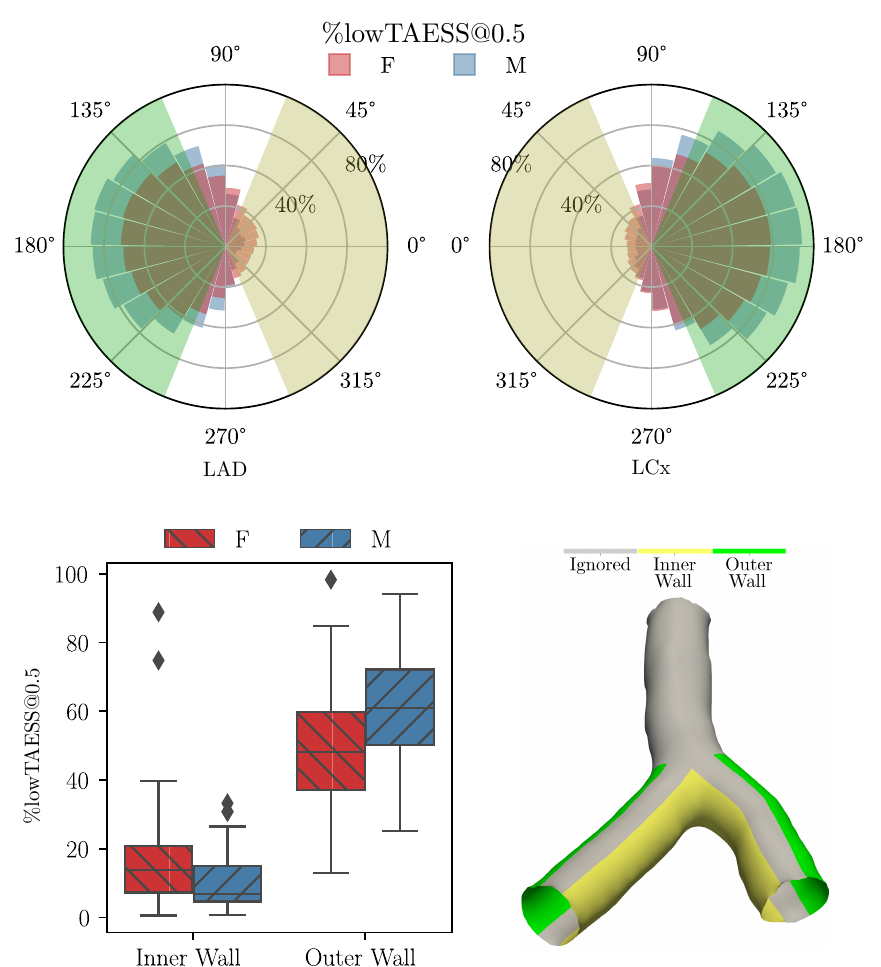}
\caption{Radial difference by sex in \%lowTAESS@0.5 proximal to the left main carina: Percentage of \%lowTAESS@0.5 for females (red) and males (blue) based on radial location for both the LAD (top left) and LCx (top right) daughter branches. Differences in the spatial distribution of \%lowTAESS@0.5 shown as a box plot (bottom left) for the inner (yellow) and outer (green) wall regions (bottom right), again for females (red) and males (blue), here visualised for a 135\textdegree region (bottom right)). As shown in the Supplementary Material the choice of region size did not have a significant effect.}
\label{fig:cfd-hemodifferences}
\end{figure}

Whilst the overall percentage of vascular surface exposure to \%lowTAESS@0.5 was similar between males and females (M 26.8\% vs F 25.0\%, p = 0.274), an interesting difference in spatial distributions appeared whereby \%lowTAESS@0.5 was shifted towards the flow divider and inner vessel walls in female patients. Specifically, when defining the inner and outer vessel regions (using 90, 135 or 180\textdegree wedges centred along the bifurcation plane, Fig. \ref{fig:cfd-hemodifferences}), the percentage of the inner wall exposed to \%lowTAESS@0.5 was significantly higher in females (M 10.6\% vs F 16.8\%, p=0.004*). Similarly, for the outer walls, \%lowTAESS@0.5 was higher in males (M 59.9\% vs F 47.8\%, p\textless0.001*). See the Supplementary Material for details (Fig. \ref{fig:innerouter}). 

Differences in diameter (R$^2$=0.46, p\textless0.001*) and mean curvature between the two groups (R$^2$=0.37, p\textless0.001* ) appear to drive this regional shift in \%lowTAESS@0.5 when using linear regression models between the shape features and the difference in \%lowTAESS@0.5 for the inner/outer wall distribution. Specifically, smaller vessel diameters and higher overall bifurcation curvature shift adverse \%lowTAESS@0.5 distribution towards the inner proximal daughter branch vessel walls.

Men showed higher adverse values in both \%highOSI@0.1 (M 4.57\% vs F 2.34\%, p \textless 0.001*), and \%highRRT@4.17 (M 6.6\% vs F 4.24\%, p = 0.005*), which infers higher disturbed and oscillatory flow conditions in males compared to females (see Table \ref{tbl:cfd-hemo} and Fig. \ref{fig:cfd-hemodifferences}).

All reported differences in \%highOSI@0.1, \%highRRT@4.17, and spatial distribution of \%lowTAESS@0.5 remained statistically significant when adjusted for age, BMI, BSA, smoking history, and hypertension in a multivariate model.

\begin{table}[h]
\centering
\caption{Haemodynamic differences between suspected CAD males and females left main coronary bifurcations without significant stenosis.}
\label{tbl:cfd-hemo}
\rowcolors{2}{gray!25}{white}
\begin{tabular}{lS[table-format=2.2(2)]S[table-format=2.2(2)]S[table-format=2.3]cc}
&&&&\textbf{Difference} &\\
\textbf{Metric} & {\textbf{Male}} & \textbf{Female} & \textbf{p-value} & \textbf{M – F} & \textbf{95\% CI} \\\hline
\textbf{\%lowTAESS@0.5\%} &26.83(8.00) &	25.01(9.90) &	0.274 &	1.818  & {[}-1.46  5.10{]}   \\
\textbf{\%lowTAESS@0.5\%, Inner wall}&10.71(8.76) &	16.78(14.47) &	0.004* &	-6.07& {[}-10.21  -1.93{]}   \\
\textbf{\%lowTAESS@0.5\%, Outer wall} & 59.88(17.21) & 47.87(16.15)&	\textless0.001* &	12.01  & {[}5.61 18.42{]} \\
\textbf{\%highOSI@0.1\%} &4.57(2.80) & 2.35(1.68) & \textless0.001* &	2.22 & {[}1.27 3.18{]}   \\
\textbf{\%highRRT@4.17\%} &6.06(3.56) & 4.24(2.57) &	0.005* &	1.82 & {[}0.58 3.07{]}
\end{tabular}
\end{table}

\begin{figure}[h]
\centering
\includegraphics[width=0.5\linewidth]{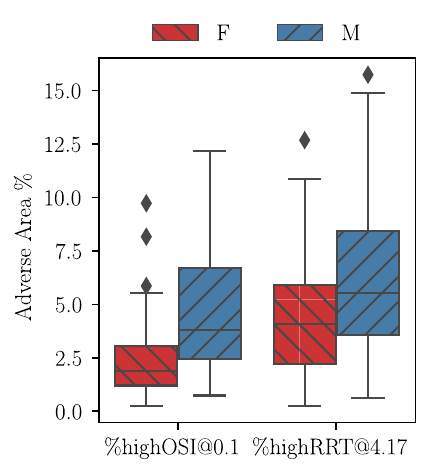}
\caption{Sex differences in left main OSI and RRT: Differences in normalised percentage vessel area coverage of adversely high Oscillatory Shear Index (\%highOSI@0.1, left) and Relative Resident Time (\%highRRT@4.17, right) haemodynamic metrics in females (red) and males (blue).}
\label{fig:cfd-OSIRRTdifferences}
\end{figure}

\section{Discussion}
This work quantifies the differences between male and female patients undergoing CTCA imaging for suspected CAD with no significant coronary artery stenosis, analysing anatomical and haemodynamic metrics. 

Shape characteristics, such as the average vessel diameter\cite{huo2012diameter, papakonstantinou2013sex} and curvature\cite{shen2021secondary,chiastra2017healthy} of coronary arteries, have been established as potential anatomical risk markers owing to their demonstrated effect in generating local haemodynamics with clinically adverse associations. However, it should be noted that some or many of these characteristics may have interconnected effects, like the bifurcation angle, which was demonstrated to only generate adverse haemodynamics in combination with other anatomical characteristics \cite{beier_ImpactBifurcationAngle_2016}.

We found vessel diameters to be larger in males compared to females, even after adjusting for BMI and body surface area, which is in line with previously reported results \cite{ilayperuma_SexualDifferencesDiameter_2011}.  Additionally, it should be noted that our prior work \cite{medrano2016computational} reported differences in Angle B between males and females when cases with intermediate arteries were included. Here, after excluding intermediate arteries,  no statistically significant difference in Angle B was found.

Tortuosity is well known to be greater in females than males \cite{li_ClinicalImplicationCoronary_2011,eleid_CoronaryArteryTortuosity_2014}, although it has often been measured using unreliable metrics in literature and thus previously reported findings using the `tortuosity index (rather than the recommended mean curvature) need to be interpreted with caution \cite{kashyap2022accuracy}. Increased arterial tortuosity is associated with both atherosclerotic coronary disease and non-atherosclerotic Spontaneous Coronary Artery Dissection (SCAD), a less common condition presenting at a younger age and affecting women far more often than men \cite{saw2014nonatherosclerotic}. It has been proposed that increased arterial curvature would cause an increase in shear stress, which may weaken the vascular wall and ultimately lead to SCAD \cite{eleid_CoronaryArteryTortuosity_2014, ciurica_ArterialTortuosity_2019}. Recent work also linked SCAD to additional measures describing regional extreme TAESS zones \cite{carpenter2020review, candreva2023spontaneous}). Previous studies found sex-specific differences in plaque size, and also differences in ESS when stratified by age \cite{wentzel2022sex}. However, this study did not consider the radial position of plaques, and thus, a direct comparison is difficult. Further, our cohort did not allow a meaningful analysis in terms of age, although it should be noted that the mean age differed statistically between the two groups, with males being approximately 4 years younger than females. Previous studies of carotid arteries \cite{tajik2012asymmetrical} did not show a difference in the radial distribution of plaques between male and female patients. To our knowledge, no previous study has explored sex-specific radial haemodynamics in coronaries before our work presented here. 

Similarly, the effect of the significant sex-specific difference in inflow angle and diameter, as found here, their haemodynamic impact, and, thus, their potential clinical association has not been explored or reported before and warrants future studies.  

In women, coronary artery disease often manifests as a non-obstructive disorder \cite{chaudhary2019gender}, and is more challenging to diagnose, indicating the need for sex-specific patient care \cite{pilote2007comprehensive}. Non-obstructive CAD correlates to severe endothelial dysfunction through low ESS \cite{kumar_LowCoronaryWall_2018}. Whilst \%lowTAESS@0.5 was not different between males and females, the spatial distribution differed. This may result in different locations of plaque development, and may also affect its progression by inducing different haemodynamics. This would not be observable by today's standard imaging modalities, which are usually portrayed longitudinally, and too limited in resolution to observe nuanced axial differences.

We also observed higher oscillatory conditions in males, a predictor of neointimal thickening, which is the initial stage of atherosclerosis development \cite{kolodgie2007pathologic}. This may indicate that males are haemodynamically more prone to obstructive lesions compared to females. Since the males in this cohort also had no significant stenosis, an equivalent study in males versus females should be undertaken in those with obstructive disease to investigate this further.

The inclusion of sex as a CAD risk factor is clinically relevant. Males develop atherosclerotic CAD at a younger age than females, and females are more prone to non-obstructive CAD and non-atherosclerotic coronary artery disease, such as SCAD. Hence, disease risk stratification must be viewed differently in the context of the patient’s sex regarding their clinical diagnosis, management, and outcomes \cite{eastwood_GenderDifferencesCoronary_2005}. There have not been any prospective longitudinal studies on the implications of \%lowTAESS@0.5 on the development of CAD, which would inform the understanding of the onset and progression of atherosclerosis. As mentioned, certain conditions like SCAD may be due, at least in part, to greater vessel curvature in females. Hypertensive women have a high risk of non-obstructive CAD. Moreover, hypertension often starts around menopausal in females, and even mild hypertension (140/90 mmHg) causes endothelial dysfunction, leading to disease \cite{maas_GenderDifferencesCoronary_2010}. 

The new insights from evaluating potential geometric risk factors and their local haemodynamic characteristics may be critical to understanding CAD risk factors better in both sexes. Significantly, women may benefit from this due to commonly atypical and under-recognised presentations. Improved prediction and early prevention may be the key to improving outcomes. 

Future studies should include an obstructive disease cohort, and may also elaborate further on other compounding factors, such as hypertension, diabetes, and pre- and post-menopausal effects, to inform longitudinal risk stratification effective in both sexes. 

\section{Conclusion}
This work has shown significant differences in the anatomical features and resulting haemodynamic profiles of LM bifurcations in males and females with suspected CAD but non-significant stenosis. Namely, \%lowTAESS@0.5 was more commonly found in the inner proximal bifurcation walls in females and outer walls in males. Males showed significantly higher \%highOSI@0.1, likely due to larger vessel size, fewer curvatures and more accurate inflow angles than females. These findings may have future implications for our understanding of sex-specific CAD development and inform more personalised assessment and treatment considerations. Further studies are warranted to investigate underlying mechanisms that may lead to sex-specific screening and therapeutic strategies and, thus, improved outcomes for women and men with CAD.

\section{Acknowledgements}
SB acknowledges Intra for assistance in collection of the dataset and the Auckland Academic Health Alliance (AAHA) and the Auckland Medical Research Foundation (AMRF) for
their financial support and endorsement.
This research was undertaken with the assistance of computational resources from the National Computational Infrastructure (NCI), which is supported by the Australian Government, and the computational cluster Katana supported by Research Technology Services at UNSW Sydney.

\bibliography{refs}

\newpage
\appendix
\section*{Supplementary Material}
\label{sec:supmat}
\renewcommand{\thefigure}{A.\arabic{figure}}
\setcounter{figure}{0}    
\begin{figure}[t]
\centering
\includegraphics[width=0.9\linewidth]{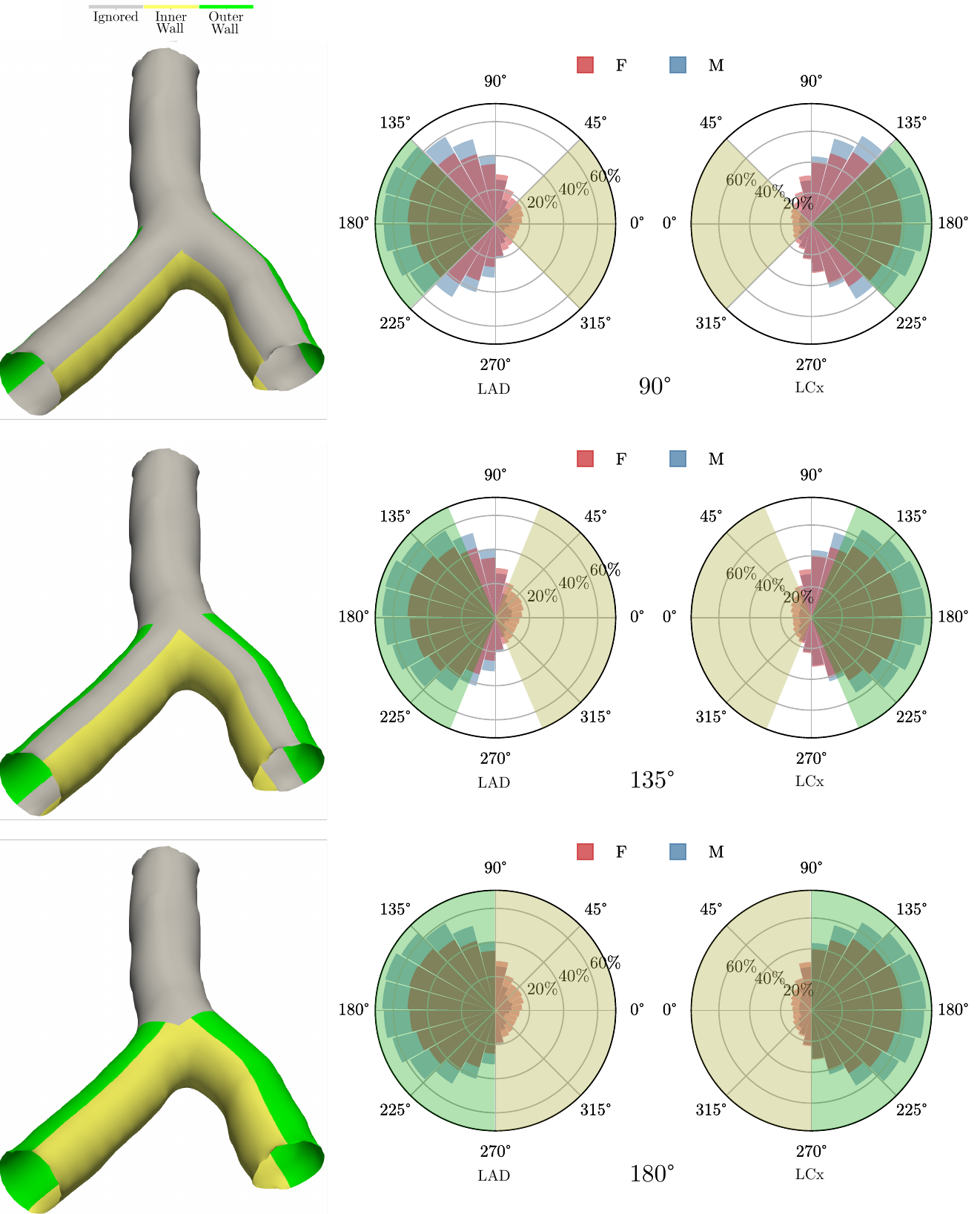}
\caption{Inner and outer wall classifications for sex-specific \%lowTAESS@0.5 radial distribution: Percentage of \%lowTAESS@0.5 for females (red) and males (blue) based on radial location for both the Left Anterior Descending LAD (middle) and Left Circumflex LCx (right) daughter branch. Sensitivity analysis based on different region sizes of 90, 135, and 180\textdegree wedges (top to bottom) for inner (yellow) and outer (green) regions showed no difference in the results obtained.}
\label{fig:innerouter}
\end{figure}
\end{document}